\documentclass[12pt,preprint]{aastex}

\slugcomment{{\it Submitted to} Ap.~J.~Lett., {\it October 16, 2007.} \\ 
{\it Re-submitted January 15, 2008, Available at} arXiv:0710.3399}

\shorttitle{How Reconnection Regulates Coronal Conditions}            
\shortauthors{Cassak, Mullan and Shay}

\begin{document}

\title{From Solar and Stellar Flares to Coronal Heating: \\
  Theory and Observations of How Magnetic \\ Reconnection 
  Regulates Coronal Conditions}

\author{P.~A.~Cassak\altaffilmark{1}, D.~J.~Mullan\altaffilmark{1}
and M.~A.~Shay\altaffilmark{1}}

\altaffiltext{1}{Department of Physics and Astronomy and Bartol
Research Institute, University of Delaware, 217 Sharp Laboratory,
Newark, DE 19716, USA; pcassak@udel.edu, mullan@udel.edu,
shay@physics.udel.edu}

\begin{abstract}
There is currently no explanation of why the corona has the
temperature and density it has.  We present a model which explains how
the dynamics of magnetic reconnection regulates the conditions in the
corona.  A bifurcation in magnetic reconnection at a critical state
enforces an upper bound on the coronal temperature for a given
density.  We present observational evidence from 107 flares in 37
sun-like stars that stellar coronae are near this critical state.  The
model may be important to self-organized criticality models of the
solar corona.
\end{abstract}





\section{INTRODUCTION}

Dynamics in the solar corona takes on a wide range of forms.  On one
hand, the corona is the setting for the most violent eruptions in the
solar system: solar flares and coronal mass ejections
\citep{Aschwanden01}.  On the other, coronal heating makes the corona
almost a thousand times hotter than the photosphere, even in the quiet
sun \citep{Klimchuk06}.  \citet{Parker83,Parker88} unified these two
phenomena by proposing that micro- and nano-flares, less energetic
cousins of eruptive flares, heat the corona.  This model gained
credence from studies showing that solar flares exhibit power law
statistics \citep{Lin84,Dennis85,Crosby93,Feldman97,Wheatland00,
Nita02,Paczuski05} over a wide range of scales for many quantities.
[See \citet{Charbonneau01} for a review.]  In addition, stellar flares
have similar light curves to solar flares \citep{Gershberg05} and also
exhibit power law statistics \citep{Collura88,Shakhovskaya89,
Audard00}, suggesting that the physics of the solar corona is generic
to sun-like stars.

Coronal dynamics remains an active research area \citep{Hudson91,
Georgoulis98,Shibata02,Hughes03}.  Details of the eruption process
including how magnetic energy is stored, how eruptions onset, and how
the stored energy is converted to other forms are still open
questions.  In addition, while micro- and nano-flares are believed to
be a major contributor to coronal heating, the authors know of no
theory which explains why the coronal temperature and density have the
values they have, as opposed to larger or smaller values.

In this paper, we propose that the condition of the corona is
regulated by magnetic reconnection \citep{Cassak06b}, a dynamical
process which converts magnetic energy into kinetic energy and heat
and energizes particles.  Magnetic energy is stored during collisional
(slow) reconnection, which has been shown to drive the coronal plasma
toward lower collisionality \citep{Cassak06}.  If the plasma becomes
marginally collisionless, a bifurcation in the underlying dynamics of
reconnection occurs \citep{Cassak07c}.  This bifurcation, which occurs
when two length scales $\delta_{SP}$ and $\rho_{i}$ (to be defined
below) are comparable, catastrophically initiates fast (Hall)
reconnection, releasing the stored energy in the form of an eruption.
The condition of marginal collisionality, therefore, sets an upper
bound on how hot the coronal plasma can be for a given density.  The
continual driving toward lower collisionality of the pre-flare corona
by slow reconnection enforces the self-organization of the corona to a
state of marginal collisionality where $\delta_{SP} \sim \rho_{i}$.
We present finer details of this process below.  Then, we perform the
first observational test of this model using a large sample of data
from stellar flares on sun-like stars.  We find that $\delta_{SP}$ and
$\rho_{i}$ are comparable for every event in the sample, indicating
that stellar coronae do self-organize into a marginally collisional
state.

\section{THEORY}

Magnetic reconnection depends strongly on the collisionality of the
plasma.  Collisional (Sweet-Parker) reconnection \citep{Sweet58,
Parker57} is exceedingly slow.  The thickness $\delta_{SP}$ of the
Sweet-Parker diffusion region is given by \citep{Parker57}
\begin{equation}
\delta_{SP} \sim \sqrt{\frac{\eta c^{2}}{4 \pi c_{A}} L_{SP}},
\label{deltaspdef}
\end{equation}
where $c_{A} = B / (4 \pi m_{i} n)^{1/2}$ is the Alfv\'en speed, $B$
is the strength of the reconnecting magnetic field, $m_{i}$ is the ion
mass, $n$ is the density, $\eta$ is the resistivity, and $L_{SP}$ is
the length of the Sweet-Parker diffusion region in the outflow
direction.  The normalized reconnection rate $v_{in} / c_{A} \sim
\delta_{SP} / L_{SP}$ is $10^{-7}$ for coronal parameters, where
$v_{in}$ is the inflow speed.  Collisionless (Hall) reconnection has a
reconnection rate of the order of 0.1 \citep{Shay99,Birn01}, six
orders of magnitude faster than Sweet-Parker reconnection for coronal
parameters.

Recent studies \citep{Cassak05,Cassak07a} showed that the transition
from collisional to collisionless reconnection is catastrophic,
occurring when $\delta_{SP}$ becomes smaller than the ion gyroradius
$\rho_{i}$.  At this scale, magnetohydrodynamics (MHD) breaks down and
the Hall effect (absent in MHD) allows reconnection to be fast
\citep{Birn01, Rogers01}.  The transition can be described as a
bifurcation \citep{Cassak06b,Cassak07c} which takes stable equilibria
out of existence as a control parameter ($\delta_{SP} / \rho_{i}$)
varies.  The relevant gyroradius $\rho_{i}$ for anti-parallel
reconnection is the ion inertial scale $d_{i}$ \citep{Cassak05},
\begin{equation}
d_{i} = \frac{c_{A}}{\Omega_{ci}} = \frac{c}{\omega_{pi}} =
\sqrt{\frac{m_{i} c^{2}}{4 \pi n e^{2}}}, \label{didef}
\end{equation}
where $\Omega_{ci}$ is the ion cyclotron frequency, $\omega_{pi}$ is
the ion plasma frequency, and $e$ is the ion charge.  For reconnection
with a (guide) magnetic field along the current sheet, the relevant
gyroradius becomes $\rho_{s} = c_{s} / \Omega_{ci}$, where $c_{s}$ is
the sound speed \citep{Cassak07a}.

To see how magnetic reconnection self-organizes the corona, consider
an active region.  Before an eruption, the plasma cannot be
collisionless: if it were, the stored magnetic energy would be rapidly
released by Hall reconnection.  Therefore, the pre-flare active region
must be collisional.  Since (collisional) Sweet-Parker reconnection is
exceedingly slow, magnetic energy can be stored.  [By ``collisional'',
we mean $\delta_{SP} > d_{i}$, which, using eqs.~(\ref{deltaspdef})
and (\ref{didef}), is equivalent to $\nu_{ie} > c_{A} / L_{SP},$ where
$\nu_{ie} = \eta n e^{2} / m_{i}$ is the ion-electron collision
frequency.  Therefore, reconnection is collisional when the ion
transit time along the Sweet-Parker diffusion region is longer than
the ion-electron collision time.  See also \citet{Uzdensky07b}.]

When Sweet-Parker reconnection begins in the corona, {\it e.g.,} as a
result of two coronal flux tubes coming together, the reconnecting
magnetic field $B$ is initially much weaker than the strong asymptotic
magnetic field in the core of the flux tube, {\it i.e.,} the
reconnection is embedded within a wider current sheet.  From
equation~(\ref{deltaspdef}), the thickness of the diffusion region
will be relatively wide, so that $\delta_{SP} \gg \rho_{i}$.  It was
shown \citep{Cassak06} that embedded Sweet-Parker reconnection
spontaneously self-drives the current sheet to thinner scales, even
without external forcing.  This is because the reconnection inflow
convects stronger magnetic fields into the diffusion region, which
causes $\delta_{SP}$ to decrease [see eq.~(\ref{deltaspdef})].  Thus,
the reconnection process itself self-drives the system toward lower
collisionality.

If the asymptotic field is strong enough so that $\delta_{SP} \sim
\rho_{i}$, {\it i.e.,} the system becomes marginally collisionless,
then a bifurcation causes Hall reconnection to begin, eruptively
releasing the stored energy.  [We note in passing that if $B$ in an
active region is not strong enough to ever satisfy $\delta_{SP} \sim
d_{i}$ for a given density and temperature then no eruption occurs,
potentially providing an observational constraint on which active
regions erupt and which do not.]  After the eruption, the corona
returns to a collisional state, and the process begins again.  The
continual self-driving of the corona toward lower collisionality keeps
coronal parameters near the critical condition where the bifurcation
occurs ($\delta_{SP} \sim \rho_{i}$).

We propose that this process regulates the temperature of the corona.
If the temperature $T$ of the corona is larger than the critical
value, the Spitzer resistivity $\eta$ is smaller (since $\eta \propto
T^{-3/2}$).  From equation~(\ref{deltaspdef}), a smaller $\eta$ allows
a smaller $B$ to initiate an eruption.  As such, less magnetic energy
is stored and released, and the corona cools.  Alternately, if the
corona is cooler than the critical value, a larger magnetic field is
required to set off an eruption.  More magnetic energy is stored and
released, increasing the temperature.  Either way, the temperature is
driven back to the critical value.  \citet{Uzdensky06,Uzdensky07,
Uzdensky07b} independently proposed a similar model based on the
density, which we discuss in Sec.~\ref{sec-disc}.

\section{OBSERVATIONAL DATA AND RESULTS}
\label{sec-spandhall}

Observational verification of this model entails confirming that
$\delta_{SP}$ and $\rho_{i}$ are comparable at fast reconnection
onset.  Laboratory experiments \citep{Ren05,Egedal07} are consistent
with this condition, but direct observations are impossible because
the length scales are not resolvable.  Indirect verification is
possible by estimating $\delta_{SP}$ and $\rho_{i}$ for coronal
parameters.  For a solar active region, one finds both length scales
to be a few meters, as has been noted previously \citep{Priest00,
Uzdensky03,Bhattacharjee04,Cassak05,Cassak06,Uzdensky07b}.  An
important question is whether this agreement is indicative of a
general mechanism or is just a coincidence for solar parameters.

We use data from a recent study \citep{Mullan06} which analyzed 134
eruptive flares from 44 stars of spectral type F, G, K, and M, using
the Deep Survey/Spectrometer Instrument on the {\it Extreme
Ultraviolet Explorer} ({\it EUVE}) satellite.  The data come from a
single instrument on a single satellite, so there are no spurious
variations due to different instrumental characteristics.  See
\citet{Mullan06} for a thorough discussion of the data.

The $e$-folding time $\tau_{d}$ for the flare signal to decay and the
emission measure $EM$ were extracted from flare light curves.  Because
of interruptions in {\it EUVE} data during some flares,
\citet{Mullan06} presented the $\tau_{d}$ values for each flare as a
unique value, a range of values or an upper bound.  We retain events
having a particular value or a range (using the average) and omit
events given as bounds, leaving 107 events from 37 stars.

Post-flare parameters (the temperature $T$, density $n$, minimum
magnetic field $B_{min}$, and length $L$ and cross sectional area $A$
of coronal loops) were derived \citep{Mullan06} from $\tau_{d}$ and
$EM$ using an approach due to \citet{Haisch83}, which assumes that
$\tau_{d}$ is comparable to the radiative and conductive cooling
times.  Taking $A \sim (L / 10)^{2}$, one finds \citep{Haisch83}
\begin{eqnarray}
T({\rm K})           & = & \alpha_{T} (EM)^{0.25}  \tau_{d}^{-0.25}  
\nonumber \\
n({\rm cm}^{-3}) & = & \alpha_{n} (EM)^{0.125} \tau_{d}^{-1.125} 
\label{haisch} \\
L({\rm cm})          & = & \alpha_{L} (EM)^{0.25}  \tau_{d}^{0.75} 
\nonumber
\end{eqnarray}
where $\alpha_{T} = 4 \times 10^{-5}, \alpha_{n} = 10^{9}$, and
$\alpha_{L} = 5 \times 10^{-6}$ are constants (in cgs units).  A lower
bound for the magnetic field $B_{min}$ is estimated by requiring the
magnetic pressure $B^{2} / 8 \pi$ to be at least as large as the gas
pressure $2 n k_{B} T$, where $k_{B}$ is Boltzmann's constant, to
maintain a coronal loop.  As a test of the model, \citet{Mullan06}
surveyed the literature for independent measurements of $T, n, L$, and
$B_{min}$ for the stars in their study, finding that 178 of 212
measurements were consistent with the Haisch model.  This justifies
treating the derived parameters as valid independent of the Haisch
model.

We first verify that the Haisch model gives reasonable results for
solar parameters.  Eruptive solar flares have $\tau_{d} \sim 10^{4-5}$
sec and $EM \sim 10^{49-50} {\rm cm}^{-3}$ \citep{Priest00}.  Using
these values, the Haisch model predicts post-flare parameters of $T
\sim (4-13)$ MK, $n \sim (0.3-5.6) \times 10^{10}$ ${\rm cm}^{-3}$, $L
\sim 10^{10-11}$ cm, and $B_{min} \sim (10 - 70)$ G.  Compact solar
flares have $\tau_{d} \sim 10^{3}$ sec and $EM \sim 10^{47-49}$ ${\rm
cm}^{-3}$ \citep{Priest00}.  Using these values, the Haisch model
gives $T \sim (4-13)$ MK, $n \sim (3 - 5) \times 10^{11}$ ${\rm
cm}^{-3}$, $L \sim (0.5 - 1.6) \times 10^{9}$ cm, and $B_{min} \sim
(90 - 220)$ G.  These ranges of $T, n, L,$ and $B_{min}$ are
consistent with independent empirical values obtained from images and
X-ray data for flaring loops in the sun \citep{Feldman95,Shibata02}.

To calculate $\delta_{SP}$ and $d_{i}$ from the Haisch model, we use
$B_{min}$ for the upstream magnetic field and $n$ for the density.
Sweet-Parker current sheets extend to system scales \citep{Biskamp86},
so $L_{SP}$ is on the order of the coronal loop radius $A^{1/2} \sim L
/ 10$, consistent with the Haisch model.  Finally, we use $T$ to
calculate the Spitzer resistivity \citep{Spitzer53}
\begin{equation}
\eta = \frac{16 \sqrt{\pi} e^{2} \ln{\Lambda}}{3 m_{e}}
\left(\frac{m_{e}}{2 k_{B} T}\right)^{3/2},  \label{spitzer}
\end{equation}
where $m_{e}$ is the electron mass and $\ln \Lambda = \ln [ (3 / 2
e^{3}) (k_{B}^{3} T^{3} / \pi n)^{1/2}]$ is the Coulomb logarithm.
Use of this formula is justified because the electron mean free path
($\lambda_{{\rm mfp,e}} \sim v_{th,e} / \nu_{ei} \sim 25$ km for solar
conditions, where $v_{th,e}$ is the electron thermal speed and
$\nu_{ei}$ is the electron-ion collision frequency) is small compared
to length scales in the outflow direction ($L_{SP} \sim 10^{4}$ km)
and along the current sheet ($L ~ \sim 10^{5}$ km).

\begin{figure}
\plotone{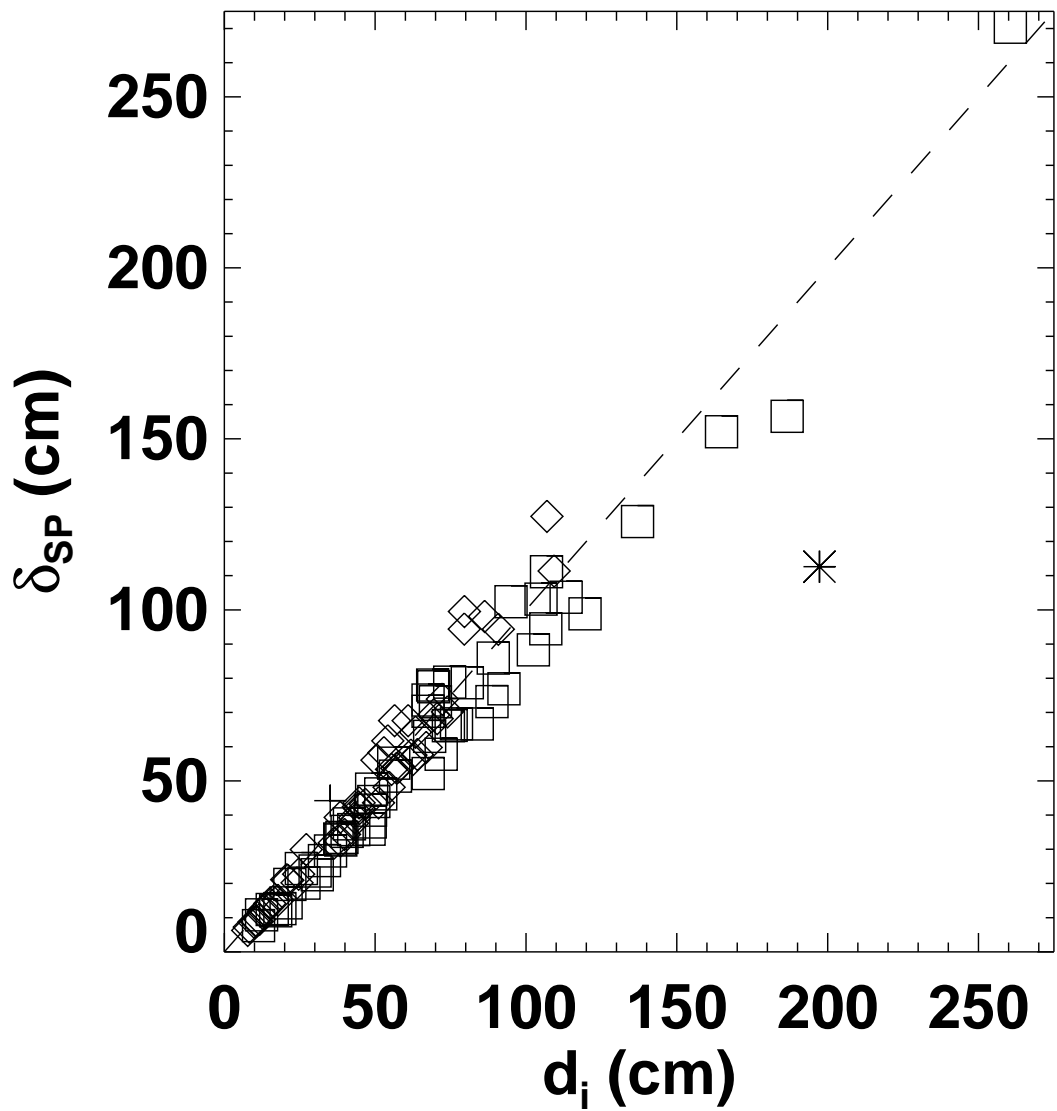}
\caption{\label{spvsdi} Sweet-Parker current layer thickness
$\delta_{SP}$ vs.~ion inertial length $d_{i} = c / \omega_{pi}$ for
the stars in the sample.  The dashed line displays their predicted
equality.  Boxes denote F, G, and K stars; diamonds denote M dwarfs.
The asterisk at $d_{i} \sim 200$ cm and the plus at $d_{i} \sim 40$ cm
denote values based on average $\tau_{d}$ and $EM$ values for eruptive
and compact solar flares, respectively \citep{Priest00}.}
\end{figure}

The result of comparing $\delta_{SP}$ to $d_{i}$ using the stellar
flare data is plotted in Fig.~\ref{spvsdi}.  Representative solar
values based on $\tau_{d} = 10^{4.5}$ sec and $EM = 10^{49.5} {\rm
cm}^{-3}$ for eruptive flares ($\delta_{SP} \sim 110$ cm and $d_{i}
\sim 200$ cm) and $\tau_{d} = 10^{3}$ sec and $EM = 10^{48} {\rm
cm}^{-3}$ for compact flares ($\delta_{SP} \sim 44$ cm and $d_{i} \sim
35$ cm) are plotted as the asterisk and plus, respectively.  A dashed
line with slope of unity is plotted.  The agreement is extremely good.
A least squares analysis gives a best fit slope of $0.98 \pm 0.02$
with a correlation coefficient of 0.981.

It is encouraging that the slope of the line in Fig.~\ref{spvsdi} is
consistent with unity.  However, there are ambiguities in the data
analysis.  For example, we used $d_{i}$ as the critical length scale,
whereas $\rho_{s}$ is more applicable to the corona (but more
difficult to estimate).  These scales differ by a factor of
$\beta_{tot}^{1/2}$, where $\beta_{tot}$ is the ratio of gas pressure
to total magnetic pressure.  If $\beta_{tot} \sim 0.1$ in the corona,
this introduces a factor of a few.  The present analysis does not
intend to distinguish between the two gyroradii; rather, the results
demonstrate that $\delta_{SP}$ is within a factor of a few of the
critical length scale $\rho_{i}$ in active stellar coronae.

A caveat of the result in Fig.~\ref{spvsdi} pertains to how the
parameters are derived in the Haisch model.  Using equations
(\ref{deltaspdef}), (\ref{didef}), and (\ref{spitzer}), using $L_{SP}
= L / 10$, and eliminating $B$ by defining the ratio of gas pressure
to magnetic pressure in the reconnecting magnetic field as
$\beta_{rec} = 2 n k_{B} T / (B^{2} / 8 \pi)$, we find $(\delta_{SP} /
d_{i})^{2} \sim (e^{4} \ln{\Lambda} / 15 k_{B}^{2}) (2 \pi m_{e}
\beta_{rec} / m_{i})^{1/2} (n L / T^{2})$.  Treating $\beta_{rec}$ as
a fixed parameter and eliminating $T, n,$ and $L$ using equation
(\ref{haisch}) gives
\begin{equation}
\left(\frac{\delta_{SP}}{d_{i}}\right)^{2} \sim \alpha
\frac{\alpha_{n} \alpha_{L}}{\alpha_{T}^{2}} \ln{\Lambda}
\sqrt{\beta_{rec}} \left(\frac{\tau_{d}}{EM}\right)^{1/8},
\label{dspodiandtaudem}
\end{equation}
where $\alpha = (e^{4} / 15 k_{B}^{2}) (2 \pi m_{e} / m_{i})^{1/2} =
1.09 \times 10^{-8} {\rm cm}^{2} {\rm K}^{2}$ is a constant.  The slow
dependence on $\tau_{d} / EM$ significantly suppresses scatter in the
observational data when evaluating $\delta_{SP} / d_{i}$.  However,
the magnitude of $\delta_{SP} / d_{i}$ is unconstrained by the Haisch
model, so the slope of the line in Fig.~\ref{spvsdi} being of order
unity is significant.  Furthermore, since the data obtained using the
Haisch model agrees with independent determinations of the same
quantities from other studies \citep{Mullan06}, it is reasonable to
assert that data obtained independently from the Haisch model would
fall close to the same line.

\begin{figure}
\plotone{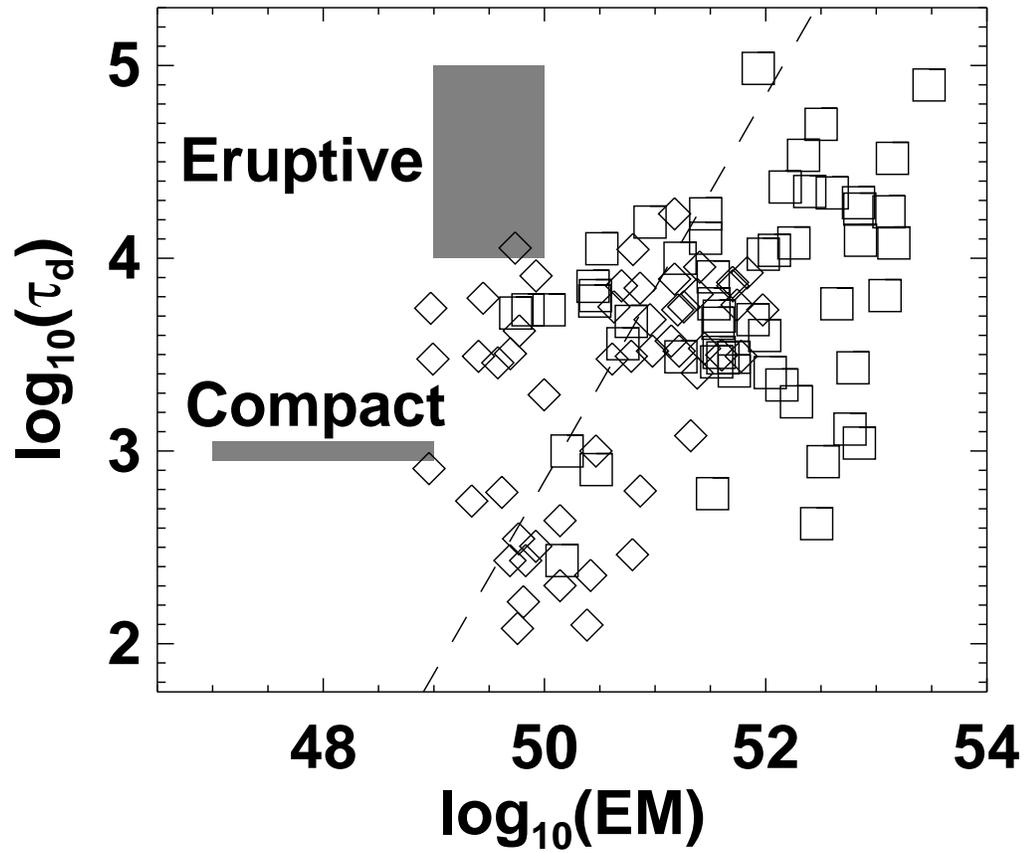}
\caption{\label{fgkstars} Decay time $\tau_{d}$ vs.~emission measure
$EM$ for the stars in the sample.  The dashed line shows the
prediction of the theory.  Boxes denote F, G, and K stars; diamonds
denote M dwarfs.  Ranges for eruptive and compact solar flares
\citep{Priest00} are shown by the gray boxes.}
\end{figure}

We can avoid suppression of the scatter in the data by solving
equation (\ref{dspodiandtaudem}) for $\tau_{d}$ and taking a logarithm
of both sides.  This yields $\log(\tau_{d}) = \log(EM) + C$, where $C
= 16 \log (\delta_{SP} / d_{i}) - 4 \log \beta_{rec} - 47$ using a
value of $\ln \Lambda \sim 22$, which is representative of the stellar
data in our study.  If $\delta_{SP} \sim d_{i}$, this predicts a
linear relationship between $\log(\tau_{d})$ and $\log(EM)$, with $C$
being the $y$-intercept.  The stellar data are plotted in
Fig.~\ref{fgkstars}.  The gray boxes show the range of values for
eruptive and compact flares on the sun \citep{Priest00}.  Assuming
$\delta_{SP} \sim d_{i}$ and taking $\beta_{rec}$ to be of order
unity, the predicted line is plotted.  While the data do not fall on a
line, the line predicted by the hypothesis that $\delta_{SP} \sim
d_{i}$ does pass through the data.  To see why this is significant,
note that if $\delta_{SP}$ was, say, $100 d_{i}$ (at $1 - 10$m, still
a very small length scale compared to coronal loop radii), then $C$
would be $-15$ instead of $-47$ and the line in Fig.~\ref{fgkstars}
would lie 32 units higher, orders of magnitude removed from the data.
The hypothesis that $\delta_{SP} \sim d_{i}$ brings significant
ordering to the data.

We note that the theory predicts $\delta_{SP} \sim d_{i}$ at flare
onset, while the Haisch model refers to post-flare conditions.
Following an eruption on the sun, temperatures typically increase by a
factor of a few \citep{Feldman95}, while the density increases due to
chromospheric evaporation by at least a factor of 10 [compare
pre-flare data \citep{Schmelz94} with post-flare data
\citep{Doschek90}].  From the relation above
equation~(\ref{dspodiandtaudem}), $\delta_{SP} / d_{i} \propto n^{1/2}
/ T$.  For an increase in $n$ by a factor of 10 and $T$ by a few,
$\delta_{SP} / d_{i}$ does not change appreciably.  Assuming this is
true for other stars, if $\delta_{SP} \sim d_{i}$ after a flare, it is
also true before a flare.

\section{DISCUSSION}
\label{sec-disc}

The data analyzed in this paper pertain to flares in sun-like stars,
but the underlying dynamics of reconnection is general.  Our model
applies equally well to micro- and nano-flares in the quiet corona.
Using values for the quiet sun of $T \sim$ 1 MK, $n \sim 10^{9}$ ${\rm
cm}^{-3}$, $B \sim $ 5 G, and $L \sim 10^{10}$ cm, we find
$\delta_{SP} \sim 770$ cm and $d_{i} \sim 720$ cm, in agreement with
the model.

The present result may have important implications for self-organized
criticality (SOC) models of the solar corona.  SOC occurs in driven,
dissipative systems when the system is driven to a critical state
where it undergoes a major reconfiguration \citep{Bak87}.  SOC leads
to power law statistics, which encouraged \citet{Lu91} to propose the
corona undergoes SOC.  Subsequent studies of SOC in the corona exist
\citep{Lu93,Vlahos95,Longcope00,Isliker01}, but a firm physical
foundation of the mechanism for self-driving and the physical
condition setting the critical state is often traded for the ease of
performing cellular automaton simulations [see \citet{Charbonneau01}
for a review].  The present result provides a physical mechanism for
self-driving (embedded Sweet-Parker reconnection) and the critical
state (marginal collisionality), which may provide an avenue for
developing quantitative predictions of SOC to compare with coronal
observations.

An alternate mechanism \citep{Uzdensky06,Uzdensky07,Uzdensky07b} for
heating the solar corona uses a change in density to achieve
self-regulation.  After an eruption, chromospheric evaporation
increases the coronal density, decreasing the ion gyroradius
[eq.~(\ref{didef})] and making subsequent eruptions more difficult.
The extent to which Uzdensky's and our mechanisms regulate coronal
heating is an open question.

The present model assumes that Sweet-Parker scaling is appropriate for
thin current sheets of large extent.  Long current sheets are known to
fragment due to secondary instabilities, but the effect of this on the
reconnection rate is unknown.  Verification of the present model would
entail testing whether Sweet-Parker reconnection in extended current
sheets remains much slower than Hall reconnection.  [See
\citet{Uzdensky07b} for further discussion of this point as well as
other future research directions.]

The authors thank J.~F.~Drake, A.~Klimas, E.~Ott, S.~Owocki, P.~So and
D.~Uzdensky for helpful conversations.  This work was supported in
part by the Delaware Space Grant.

\clearpage


\begin{thebibliography}{56}
\expandafter\ifx\csname natexlab\endcsname\relax\def\natexlab#1{#1}\fi

\bibitem[{Aschwanden {et~al.}(2001)Aschwanden, Poland, \& Rabin}]{Aschwanden01}
Aschwanden, M.~J., Poland, A.~I., \& Rabin, D.~M. 2001, ARA\&A, 39, 175

\bibitem[{Audard {et~al.}(2000)Audard, Gudel, Drake, \& Kashyap}]{Audard00}
Audard, M., Gudel, M., Drake, J.~J., \& Kashyap, V.~L. 2000, Ap.~J., 541, 396

\bibitem[{Bak {et~al.}(1987)Bak, Tang, \& Wiesenfeld}]{Bak87}
Bak, P., Tang, C., \& Wiesenfeld, K. 1987, Phys.~Rev.~Lett., 59, 381

\bibitem[{Bhattacharjee(2004)}]{Bhattacharjee04}
Bhattacharjee, A. 2004, Annu.~Rev.~Astron.~Astrophys., 42, 365

\bibitem[{Birn {et~al.}(2001)Birn, Drake, Shay, Rogers, Denton, Hesse,
  Kuznetsova, Ma, Bhattacharjee, Otto, \& Pritchett}]{Birn01}
Birn, J., Drake, J.~F., Shay, M.~A., Rogers, B.~N., Denton, R.~E., Hesse, M.,
  Kuznetsova, M., Ma, Z.~W., Bhattacharjee, A., Otto, A., \& Pritchett, P.~L.
  2001, J. Geophys. Res., 106, 3715

\bibitem[{Biskamp(1986)}]{Biskamp86}
Biskamp, D. 1986, Phys. Fluids, 29, 1520

\bibitem[{Cassak(2006)}]{Cassak06b}
Cassak, P.~A. 2006, PhD thesis, University of Maryland,
  www.physics.udel.edu/\~{}pcassak/cassakthesis.pdf

\bibitem[{Cassak {et~al.}(2006)Cassak, Drake, \& Shay}]{Cassak06}
Cassak, P.~A., Drake, J.~F., \& Shay, M.~A. 2006, Ap.~J., 644, L145

\bibitem[{Cassak {et~al.}(2007{\natexlab{a}})Cassak, Drake, \&
  Shay}]{Cassak07a}
---. 2007{\natexlab{a}}, Phys.~Plasmas, 14, 054502

\bibitem[{Cassak {et~al.}(2007{\natexlab{b}})Cassak, Drake, Shay, \&
  Eckhardt}]{Cassak07c}
Cassak, P.~A., Drake, J.~F., Shay, M.~A., \& Eckhardt, B. 2007{\natexlab{b}},
  Phys.~Rev.~Lett., 98, 215001

\bibitem[{Cassak {et~al.}(2005)Cassak, Shay, \& Drake}]{Cassak05}
Cassak, P.~A., Shay, M.~A., \& Drake, J.~F. 2005, Phys.~Rev.~Lett., 95, 235002

\bibitem[{Charbonneau {et~al.}(2001)Charbonneau, McIntosh, Liu, \&
  Bogdan}]{Charbonneau01}
Charbonneau, P., McIntosh, S.~W., Liu, H.-L., \& Bogdan, T.~J. 2001, Solar
  Phys., 203, 321

\bibitem[{Collura {et~al.}(1988)Collura, Pasquini, \& Schmitt}]{Collura88}
Collura, A., Pasquini, L., \& Schmitt, J.~H.~M.~W. 1988, Astron.~Astrophys.,
  205, 197

\bibitem[{Crosby {et~al.}(1993)Crosby, Aschwanden, \& Dennis}]{Crosby93}
Crosby, N.~B., Aschwanden, M.~J., \& Dennis, B.~R. 1993, Solar Phys., 143, 275

\bibitem[{Dennis(1985)}]{Dennis85}
Dennis, B.~R. 1985, Solar Phys., 100, 465

\bibitem[{Doschek(1990)}]{Doschek90}
Doschek, G.~A. 1990, Ap.~J.~Supp.~Ser., 73, 117

\bibitem[{Egedal {et~al.}(2007)Egedal, Fox, Katz, Porkolab, Reim, \&
  Zhang}]{Egedal07}
Egedal, J., Fox, W., Katz, N., Porkolab, M., Reim, K., \& Zhang, E. 2007,
  Phys.~Rev.~Lett., 98, 015003

\bibitem[{Feldman {et~al.}(1997)Feldman, Doschek, \& Klimchuk}]{Feldman97}
Feldman, U., Doschek, G.~A., \& Klimchuk, J.~A. 1997, Ap.~J., 474, 511

\bibitem[{Feldman {et~al.}(1995)Feldman, Laming, \& Doschek}]{Feldman95}
Feldman, U., Laming, J.~M., \& Doschek, G.~A. 1995, Ap.~J., 451, L79

\bibitem[{Georgoulis {et~al.}(1998)Georgoulis, Velli, \&
  Einaudi}]{Georgoulis98}
Georgoulis, M.~K., Velli, M., \& Einaudi, G. 1998, Ap.~J., 497, 957

\bibitem[{Gershberg(2005)}]{Gershberg05}
Gershberg, R.~E. 2005, Solar-Type Activity in Main-Sequence Stars (Berlin:
  Springer)

\bibitem[{Haisch(1983)}]{Haisch83}
Haisch, B.~M. 1983, in IAU Colloq.~71, Activity in Red-Dwarf Stars, ed. P.~B.
  Byrne \& M.~Rodono (Dordrecht: Reidel), 255

\bibitem[{Hudson(1991)}]{Hudson91}
Hudson, H.~S. 1991, Solar Phys., 133, 357

\bibitem[{Hughes {et~al.}(2003)Hughes, Paczuski, Dendy, Helander, \&
  McClements}]{Hughes03}
Hughes, D., Paczuski, M., Dendy, R.~O., Helander, P., \& McClements, K.~G.
  2003, Phys.~Rev.~Lett., 90, 131101

\bibitem[{Isliker {et~al.}(2001)Isliker, Anastasiadis, \& Vlahos}]{Isliker01}
Isliker, H., Anastasiadis, A., \& Vlahos, L. 2001, Astron.~Astrophys., 377,
  1068

\bibitem[{Klimchuk(2006)}]{Klimchuk06}
Klimchuk, J.~A. 2006, Solar Phys., 234, 41

\bibitem[{Lin {et~al.}(1984)Lin, Schwartz, Kane, Pelling, \& Hurley}]{Lin84}
Lin, R.~P., Schwartz, R.~A., Kane, S.~R., Pelling, R.~M., \& Hurley, K.~C.
  1984, Ap.~J., 283, 421

\bibitem[{Longcope \& Noonan(2000)}]{Longcope00}
Longcope, D.~W. \& Noonan, E.~J. 2000, Astrophys.~J., 542, 1088

\bibitem[{Lu \& Hamilton(1991)}]{Lu91}
Lu, E.~T. \& Hamilton, R.~J. 1991, Ap.~J., 380, L89

\bibitem[{Lu {et~al.}(1993)Lu, Hamilton, McTiernan, \& Bromund}]{Lu93}
Lu, E.~T., Hamilton, R.~J., McTiernan, J.~M., \& Bromund, K.~R. 1993, Ap.~J.,
  412, 841

\bibitem[{Mullan {et~al.}(2006)Mullan, Mathioudakis, Bloomfield, \&
  Christian}]{Mullan06}
Mullan, D.~J., Mathioudakis, M., Bloomfield, D.~S., \& Christian, D.~J. 2006,
  Ap.~J.~Supp.~Ser., 164, 173

\bibitem[{Nita {et~al.}(2002)Nita, Gary, Lanzerotti, \& Thomson}]{Nita02}
Nita, G.~M., Gary, D.~E., Lanzerotti, L.~J., \& Thomson, D.~J. 2002, Ap.~J.,
  570, 423

\bibitem[{Paczuski {et~al.}(2005)Paczuski, Boettcher, \& Baiesi}]{Paczuski05}
Paczuski, M., Boettcher, S., \& Baiesi, M. 2005, Phys.~Rev.~Lett., 95, 181102

\bibitem[{Parker(1957)}]{Parker57}
Parker, E.~N. 1957, J. Geophys. Res., 62, 509

\bibitem[{Parker(1983)}]{Parker83}
---. 1983, Ap.~J., 264, 642

\bibitem[{Parker(1988)}]{Parker88}
---. 1988, Ap.~J., 330, 474

\bibitem[{Priest \& Forbes(2000)}]{Priest00}
Priest, E. \& Forbes, T. 2000, Magnetic Reconnection (Cambridge University
  Press)

\bibitem[{Ren {et~al.}(2005)Ren, Yamada, Gerhardt, Ji, Kulsrud, \&
  Kuritsyn}]{Ren05}
Ren, Y., Yamada, M., Gerhardt, S., Ji, H., Kulsrud, R., \& Kuritsyn, A. 2005,
  Phys.~Rev.~Lett., 95, 005003

\bibitem[{Rogers {et~al.}(2001)Rogers, Denton, Drake, \& Shay}]{Rogers01}
Rogers, B.~N., Denton, R.~E., Drake, J.~F., \& Shay, M.~A. 2001, Phys. Rev.
  Lett., 87, 195004

\bibitem[{Schmelz {et~al.}(1994)Schmelz, Holman, Brosius, \&
  Willson}]{Schmelz94}
Schmelz, J.~T., Holman, G.~D., Brosius, J.~W., \& Willson, R.~F. 1994, Ap.~J.,
  434, 786

\bibitem[{Shakhovskaya(1989)}]{Shakhovskaya89}
Shakhovskaya, N.~I. 1989, Solar Phys., 121, 375

\bibitem[{Shay {et~al.}(1999)Shay, Drake, Rogers, \& Denton}]{Shay99}
Shay, M.~A., Drake, J.~F., Rogers, B.~N., \& Denton, R.~E. 1999, Geophys.
  Res. Lett., 26, 2163

\bibitem[{Shibata \& Yokoyama(2002)}]{Shibata02}
Shibata, K. \& Yokoyama, T. 2002, Ap.~J., 577, 422

\bibitem[{Spitzer \& H\"arm(1953)}]{Spitzer53}
Spitzer, L. \& H\"arm, R. 1953, Phys.~Rev., 89, 977

\bibitem[{Sweet(1958)}]{Sweet58}
Sweet, P.~A. 1958, in Electromagnetic Phenomena in Cosmical Physics, ed.
  B.~Lehnert (Cambridge University Press, New York), 123

\bibitem[{Uzdensky(2003)}]{Uzdensky03}
Uzdensky, D.~A. 2003, Ap.~J., 587, 450

\bibitem[{Uzdensky(2006)}]{Uzdensky06}
---. 2006, ArXiv Astrophysics e-print arXiv:0607656

\bibitem[{Uzdensky(2007{\natexlab{a}})}]{Uzdensky07}
---. 2007{\natexlab{a}}, Mem.~S.~A.~It., 75, 282

\bibitem[{Uzdensky(2007{\natexlab{b}})}]{Uzdensky07b}
---. 2007{\natexlab{b}}, Ap.~J., 671, 2139

\bibitem[{Vlahos {et~al.}(1995)Vlahos, Georgoulis, Kluiving, \&
  Paschos}]{Vlahos95}
Vlahos, L., Georgoulis, M., Kluiving, R., \& Paschos, P. 1995, Astron.
  Astrophys., 299, 897

\bibitem[{Wheatland(2000)}]{Wheatland00}
Wheatland, M.~S. 2000, Ap.~J., 536, L109

\end{thebibliography}

\clearpage

\clearpage

\end{document}